\documentclass[%
 reprint,
 superscriptaddress,
 amsmath,amssymb,
 prl,
]{revtex4-1}

\usepackage{float}
\usepackage{graphicx}
\usepackage{dcolumn}
\usepackage{bm}

\begin{document}

\preprint{APS/123-QED}

\title{Magnetic Microphase Inhomogeneity as a Thermodynamic Precursor of Ground State Phase Separation in Weakly Coupled Spin-$\frac{3}{2}$ Chains}

\author{L. Shen}
 \affiliation{Division of Synchrotron Radiation Research, Lund University, SE-22100 Lund, Sweden}
\author{E. Campillo}
 \affiliation{Division of Synchrotron Radiation Research, Lund University, SE-22100 Lund, Sweden}
\author{E. Young}
 \affiliation{School of Metallurgy and Materials, University of Birmingham, Birmingham B15 2TT, United Kingdom}
\author{C. Bulbucan}
 \affiliation{Division of Synchrotron Radiation Research, Lund University, SE-22100 Lund, Sweden}
 \author{R. Westerstr\"om}
 \affiliation{Division of Synchrotron Radiation Research, Lund University, SE-22100 Lund, Sweden}
 \author{M. Laver}
\affiliation{School of Metallurgy and Materials, University of Birmingham, Birmingham B15 2TT, United Kingdom}%
\author{P. J. Baker}
 \affiliation{ISIS Pulsed Neutron and Muon Source, STFC Rutherford Appleton Laboratory, Harwell Oxford, Oxfordshire OX11 0QX, United Kingdom} 
\author{E. Blackburn}
 \affiliation{Division of Synchrotron Radiation Research, Lund University, SE-22100 Lund, Sweden}


\begin{abstract}

$\gamma$-CoV$_{2}$O$_{6}$ is a quasi one-dimensional spin-$\frac{3}{2}$ magnet that possesses two distinct magnetic orders in the ground state with modulation vectors $k_\mathrm{1}$ = ($\frac{1}{2}$, 0, 0) and $k_\mathrm{2}$ = ($\frac{1}{4}$, 0, -$\frac{1}{4}$), respectively. Here, we use muon spin relaxation and rotation to reveal the thermodynamics of the magnetic phase separation in this compound. In the paramagnetic (PM) region, short-range correlated spin clusters emerge at $T_\mathrm{m}$ $\simeq$ 26 K at the $\it{partial}$ expense of the PM volume. Upon further cooling, we show that these emergent clusters become spatially coherent at $T_\mathrm{{N2}}$ = 7.5 K and eventually form the $k_\mathrm{2}$ order at $T^{\star}$ = 5.6 K, while the remaining PM spins are driven into the $k_\mathrm{1}$ state at $T_\mathrm{{N1}}$ = 6.6 K. These results stress magnetic microphase inhomogeneity as a thermodynamic precursor for the ground state phase separation in weakly coupled spin-$\frac{3}{2}$ chains.
\end{abstract}

\maketitle

\section{I. Introduction}

Quasi one-dimensional (1D) magnets form a fertile ground to explore magnetism in low dimensions \cite{Haldane2,Giamarchi}. Typically, materials of this class develop short-range 1D spin-spin correlations, the nature of which is determined by the strength of the coupling along the chain, at elevated temperatures. The coupling between the neighbouring chains, despite being significantly weaker than the intrachain coupling, will give rise to three-dimensional (3D) coherence when it becomes energetically relevant, upon further cooling. Recently, the exotic states of matter caused by this 1D-3D dimensional crossover have been extensively exploited in both quantum ($\it{S}$ = 1/2) \cite{Coldea,Grenier,Dupont} and classical ($\it{S}$ $>$ 1/2) \cite{Wiersche} spin systems.

The frustrated quantum many-body interactions in correlated-electron systems can lead to spatially inhomogeneous electronic or magnetic states \cite{Dagotto,Emery}. In quasi 1D magnets, magnetic phase separation is often dynamic \cite{Agrestini,Kamiya} or appears in the critical region of a first-order transition \cite{Schulenburg,Pereira} and therefore regarded as metastable. Static phase separation in the long-range ordered magnetic ground state, on the other hand, is much rarer \cite{Shen}. Moreover, thermodynamic information about the phase separation, which is essential to extracting the fundamental physics in relevant systems \cite{Dagotto,Emery}, has not been understood in any quasi 1D case so far.

The triclinic cobaltate compound $\gamma$-CoV$_{2}$O$_{6}$ ($\gamma$CVO) has weakly coupled zigzag chains of Co running along the $\it{b}$ axis; each chain is composed of two crystallographically inequivalent Co$^{2+}$ ($\it{S}$ = $\frac{3}{2}$) cations, Co\,(1) and Co\,(2), in a ratio of 1\,:\,2 \cite{Muller,Shen}. In the ground state, our neutron powder diffraction (NPD) study showed that $\gamma$CVO possesses two spatially separated magnetic order modulated by $k_\mathrm{1}$ = ($\frac{1}{2}$, 0, 0) and $k_\mathrm{2}$ = ($\frac{1}{4}$, 0, -$\frac{1}{4}$), respectively \cite{Shen,Note}. As the temperature increases, the $k_\mathrm{2}$ phase undergoes a commensurate-incommensurate transition at $T^{\star}$ = 5.6 K, accompanied by the loss of the long-range spin correlations \cite{Shen}. The $k_\mathrm{1}$ phase disappears at $T_\mathrm{{N1}}$ = 6.6 K, while the N\'eel temperature of the $k_\mathrm{2}$ phase ($T_\mathrm{{N2}}$) could not be unambiguously determined by NPD \cite{Shen}. This is due to the prevailing magnetic diffuse scattering that emerges around the $k_\mathrm{2}$ modulated Bragg positions above $T^{\star}$ and remains observable up to at least 25 K \cite{Shen}. In the same region, another investigation, using magnetometry and inelastic neutron scattering (INS), has revealed  strong ferromagnetic (FM) fluctuations in this compound \cite{Kimber}.

In a muon spin relaxation and rotation ($\mu$SR) experiment, 100\,$\%$ spin-polarized positive muons are implanted inside the sample and stop rapidly at the interstitial sites. The muons decay with a mean lifetime of $\it{\tau_{\mu}}$ = 2.2 $\mu$s, emitting positrons that are asymmetrically distributed in the forward and backward directions of the initial muon spin. This asymmetry can be used to determine the time evolution of the muon polarization, which is an extremely sensitive probe of the local magnetic environment \cite{Dalmas,Blundell} and ideally suited for the study of magnetic phase separation \cite{Brooks,Monteiro,Storchak}. 
 
In this work, we present a detailed $\mu$SR investigation on $\gamma$CVO. Upon cooling, the spatially homogeneous paramagnetic (PM) fluctuations break down (partially) at $T_\mathrm{m}$ $\simeq$ 26 K, as evidenced by the detection of two muon stopping environments that can be associated with the formation of local spin clusters. While the muon relaxation rate probing these emergent magnetic microphases diverges at 7.5 K, the one probing the PM volume does not diverge until the onset of the $k_\mathrm{1}$ phase at $T_\mathrm{N1}$ = 6.6 K, indicating a second magnetic phase transition at 7.5 K ($T_\mathrm{N2}$). By analyzing the temperature dependence of the $\mu$SR spectrum, we show that the transition at $T_\mathrm{N2}$ is intimately linked to the $k_\mathrm{2}$ phase. These results unveil a nontrivial thermodynamic pathway to the ground state magnetic phase separation in weakly coupled spin-$\frac{3}{2}$ chains. 

\section{II. Experimental Methods}

Powders of $\gamma$CVO were synthesized by the solid-state reaction method. A stoichiometric mixture of V$_{2}$O$_{5}$ (4N) and CoC$_{2}$O$_{4}$\,$\cdot$\,2H$_{2}$O (4N) were homogeneously ground in an agate mortar, pressed into pellets and annealed at 640$^{\circ}$C for 6 days. X-ray powder diffraction measurements were performed using a STOE STADI MP diffractometer (Cu K$_{\alpha1}$, $\lambda$ = 1.5406\,\AA{}) to confirm the crystallographic structure of our sample at room temperature. A tiny amount of impurity phase, identified as Co$_{2}$V$_{2}$O$_{7}$ \cite{He}, could be resolved in our Rietveld refinement (see Appendix A); its volume fraction ($<$ 0.5\,$\%$) is well below the $\mu$SR sensitivity threshold. The magnetic susceptibility measurements were performed in a Quantum Design MPMS3 Superconducting Quantum Interference Device (SQUID) magnetometer. $\mu$SR measurements were carried out on the MuSR instrument at the ISIS pulsed muon and neutron spallation source. Seven cylindrical pellets ($\sim$ 10\,mm in diameter and 1 mm in height) of $\gamma$CVO were mounted next to each other on a silver holder and placed in a helium cryostat with a base temperature about 1.5 K. 

\section{III. $\mu$SR data Modelling}

\begin{figure}[b]
	\centering
	\includegraphics[width=0.495\textwidth]{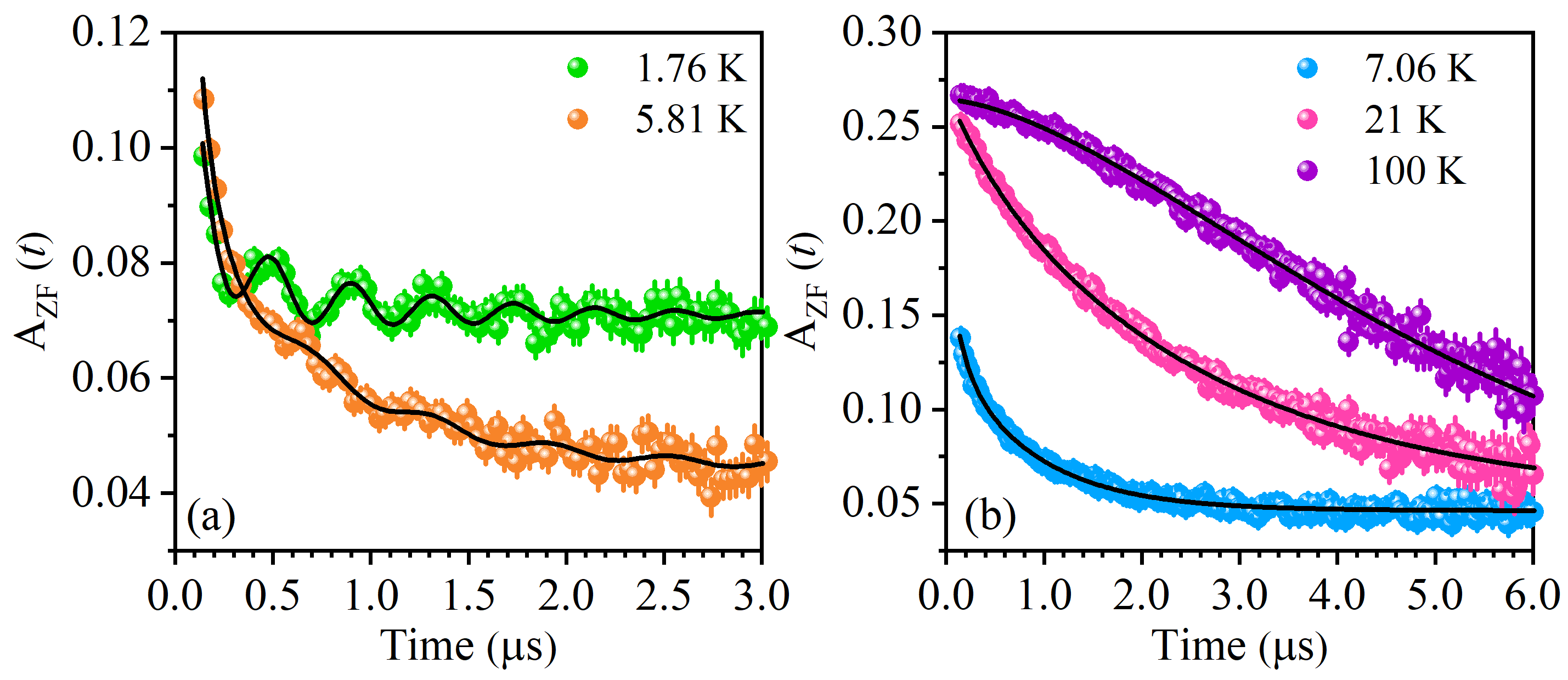}
	\caption{Selected ZF $\mu$SR spectra (a) below and (b) above $T_\mathrm{{N1}}$. The solid lines are fits using Eqs. \ref{ZF-HT} $\&$ \ref{ZF-LT}. In panel (b) the shift from a Gaussian-like to Lorentzian magnetic field distribution on cooling can be seen clearly.}
	\label{fig:1}
\end{figure}

The zero-field (ZF) $\mu$SR spectra collected above and below $T_\mathrm{{N1}}$ have been fitted by
\begin{equation}\label{ZF-HT}
    A_\mathrm{ZF}^{T > T_\mathrm{N1}} (t)=A_\mathrm{PM}e^{-(\lambda_\mathrm{PM}t)^{\beta}} + A_\mathrm{E}e^{-\lambda_\mathrm{E}t} + A_\mathrm{NM}
\end{equation}
and
\begin{equation}\label{ZF-LT}
\begin{aligned}
A_\mathrm{ZF}^{T \leq T_\mathrm{N1}} (t) = A_\mathrm{1}e^{-\lambda_\mathrm{1}t}\mathrm{cos}(2\it{\pi{}}f_{\mathrm{1}}t+\Phi_{\mathrm{1}}) + A_\mathrm{2}e^{-\lambda_\mathrm{2}t} \\
+ A_\mathrm{3}e^{-\lambda_\mathrm{3}t} + A_\mathrm{NM}.
\end{aligned}
\end{equation}

In Eq.~\ref{ZF-HT}, for $T > T_{N1}$, the first term describes the muons stopping in a homogeneous paramagnetic environment, where the exponent $\beta$ reflects the form of magnetic field distribution therein \cite{Monteiro}. The second term is only resolvable for $\it{T}$ $\leq$ $T_\mathrm{m}$ $\simeq$ 26 K; it is introduced to capture the magnetic fluctuations in the emergent spin clusters reported in Refs.~\onlinecite{Shen,Kimber}. 

In Eq.~\ref{ZF-LT}, for $T \leq T_{N1}$, the first two terms describe the coherent and incoherent muon precession about the large transverse quasistatic field generated by the magnetic long range order in the $k_\mathrm{1}$ phase \cite{Sp,Monteiro,Brooks}. The third term, to be discussed below, describes the magnetic fluctuations in the $k_\mathrm{2}$ phase.  In the modelling using Eq.~\ref{ZF-LT}, the phase $\Phi_{\mathrm{1}}$ is related to the offset between the implanted muons and detector \cite{Dalmas,Monteiro}, it has been fixed to a constant value of 5.10 degrees. 

The $A_\mathrm{NM}$ term in both equations comes from the nonmagnetic (NM) muon stopping sites, including those in the silver holder and sample. All the fits were performed on the data collected between 0.14 $\mu$s and 12.0 $\mu$s. As shown in Figure \ref{fig:1}, Eqs.~\ref{ZF-HT} $\&$ \ref{ZF-LT} reproduce the $\mu$SR spectra at the corresponding temperatures well.

\section{IV. Results}

\begin{figure}[H]
	\centering
	\includegraphics[width=0.436\textwidth]{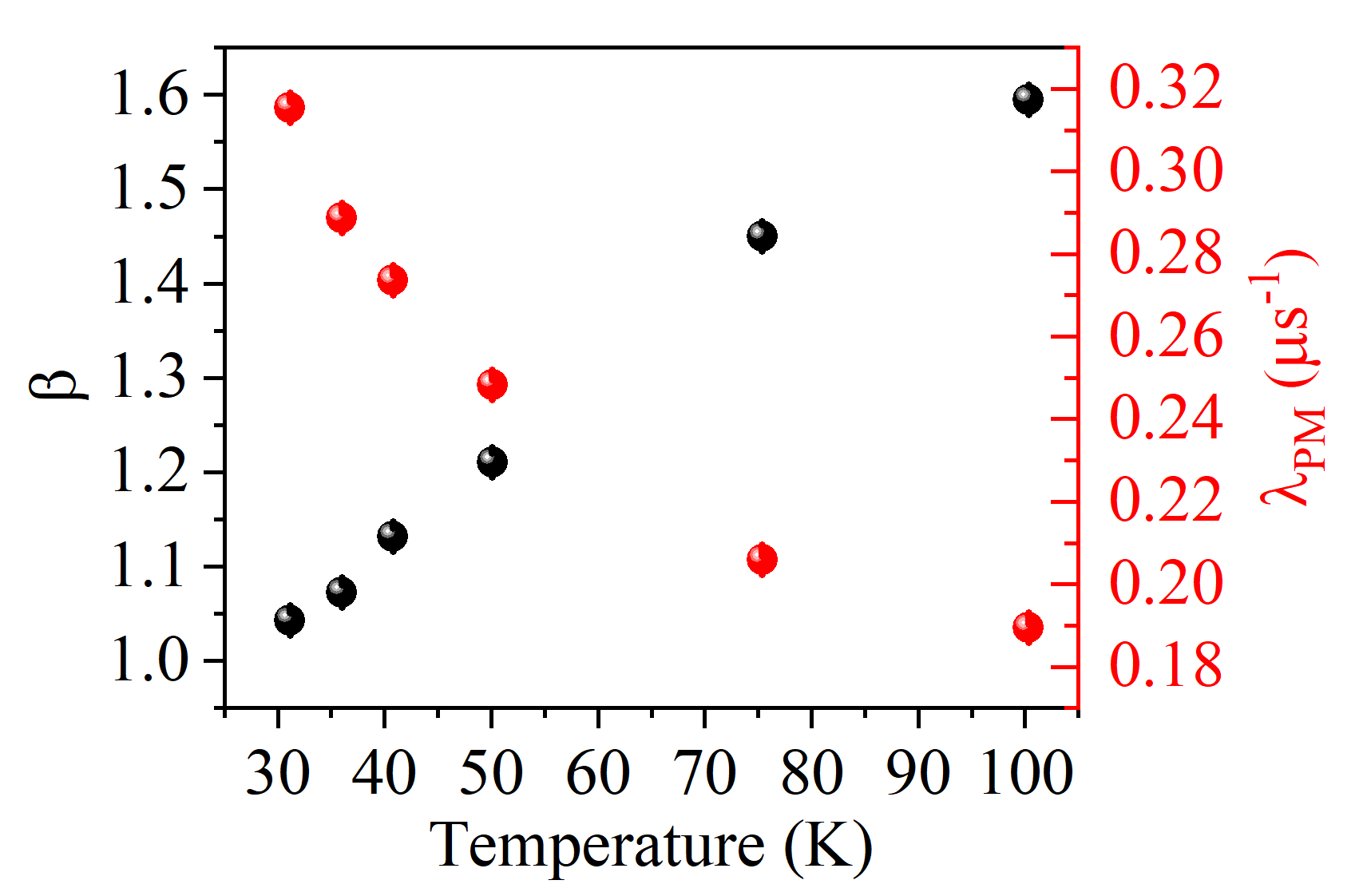}
	\caption{Temperature dependences ($\it{T}$ $>$ $T_\mathrm{m}$ $\simeq$ 26 K) of (left axis) the exponent $\beta$ in Eq.~\ref{ZF-HT} and (right axis) the relaxation rate of muons stopping in a paramagnetic environment.}
	\label{fig:2}
\end{figure}

We first discuss the muon relaxation above $T_\mathrm{m}$. In this region, we cannot resolve a finite $A_\mathrm{E}$ value, pointing to a magnetically homogeneous state. This also indicates that the first term in Eq.~\ref{ZF-HT} accounts for the whole $\gamma$CVO sample, the volume of which is therefore proportional to the initial asymmetry $A_\mathrm{PM}$ \cite{Dalmas}. $A_\mathrm{PM}$ is found to be temperature independent and equal to 0.2224(3). Moreover, $A_\mathrm{NM}$ = 0.0420(3) also does not vary in temperature within the errors; this constant amounts to the asymmetry of the silver holder probed by the muons. The temperature dependences of $\beta$ and $\lambda_\mathrm{PM}$ above $T_\mathrm{m}$ are shown in Fig.~\ref{fig:2}. At 100 K, the highest temperature measured in our study, $\beta$ is 1.59(1) and shows no sign of saturation. As the temperature decreases, $\beta$ is suppressed, reaching 1.043(7) at 31 K. This observation can be explained by a change in the nature of the dynamics probed by the muons while cooling. Moreover, since $\beta$ = 2.0 means a random magnetic field distribution, the shift towards $\beta$ = 1.0 may also suggest the build-up of the Ising-like magnetic anisotropy along the effectively isolated spin chains in this material \cite{Mannsson,Shen}. This fits with the magnetic susceptibility of our sample, which reveals deviation from the Curie-Weiss description of a completely disordered spin system well above $T_\mathrm{m}$ ($\sim$ 250 K, see Fig. \ref{fig:S2}b in Appendix C). $\lambda_\mathrm{PM}$ gradually increases upon cooling; it agrees with the typical behaviour of muons stopping in a paramagnetic environment and will be revisited in more detail later.

\begin{figure}[t]
	\centering
	\includegraphics[width=0.495\textwidth]{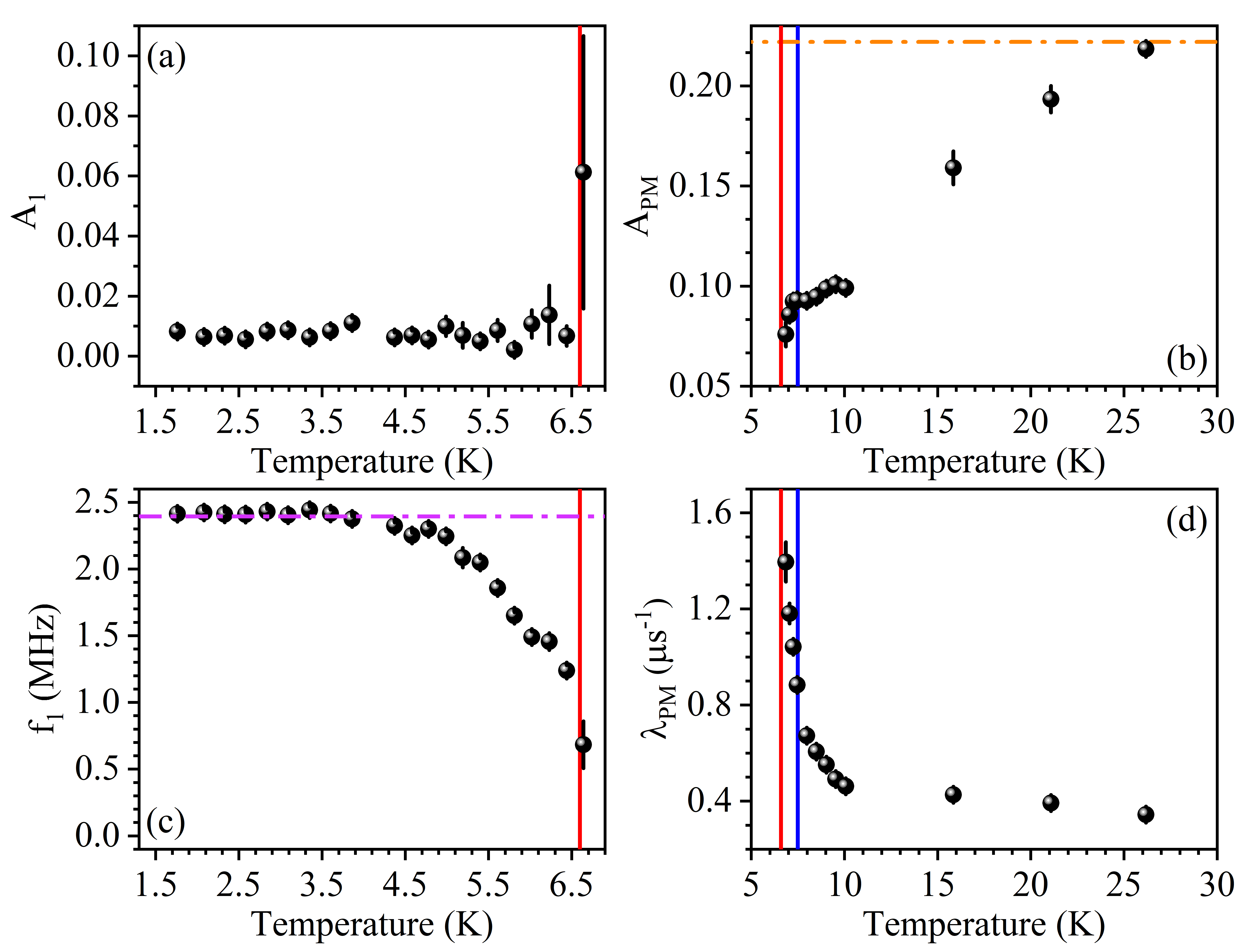}
	\caption{Temperature dependences of the (a) quasi static magnetic field asymmetry, (b) paramagnetic asymmetry, (c) oscillation frequency and (d) paramagnetic relaxation rate. These parameters are defined by the Eqs. \ref{ZF-HT} $\&$ \ref{ZF-LT} in the main text. The red and blue vertical solid lines mark the positions of $T_\mathrm{N1}$ and $T_\mathrm{N2}$. The orange horizontal dash-dot line in (b) is the total sample volume. The magenta vertical dash-dot line in (c) is guidance to eye.}
	\label{fig:3}
\end{figure}

\begin{figure}[b]
	\centering
	\includegraphics[width=0.495\textwidth]{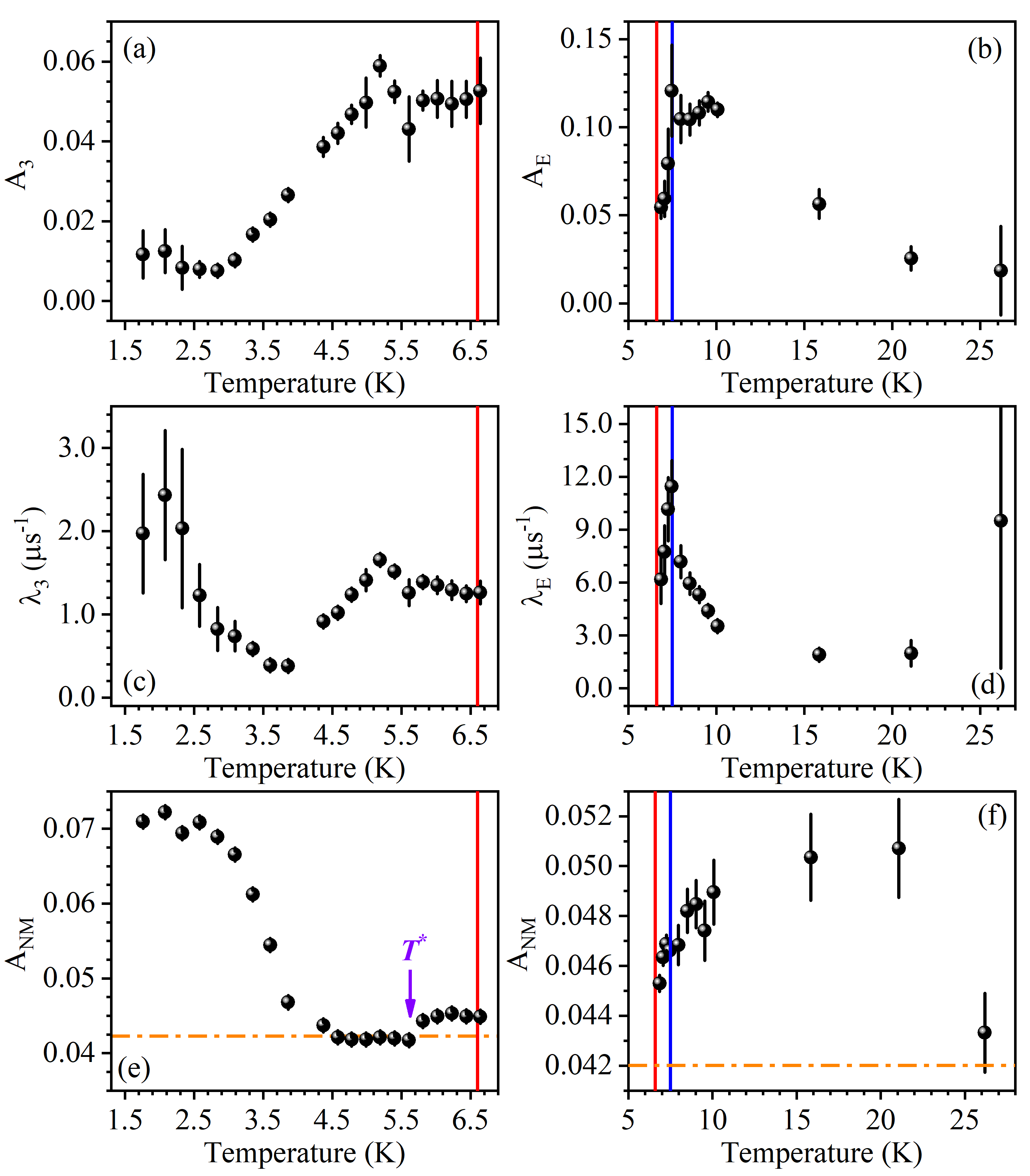}
	\caption{Temperature dependences of the (a) fluctuation volume described by the $A_{\mathrm{3}}$ term, (b) ferromagnetic cluster volume, (c) relaxation rate described by the $A_{\mathrm{3}}$ term (d) ferromagnetic cluster relaxation rate and (e-f) nonmagnetic volume. These parameters are defined by the Eqs. \ref{ZF-HT} $\&$ \ref{ZF-LT} in the main text. The red, blue and orange solid lines mark the positions of $T_\mathrm{N1}$, $T_\mathrm{N2}$ and silver background asymmetry value, respectively.}
	\label{fig:4}
\end{figure}

The exponent $\beta$ reaches 1.0 at $T_\mathrm{m}$ $\simeq$ 26 K and stops decreasing upon further cooling within our resolution. This implies that there is no distribution of relaxation times or couplings below $T_\mathrm{m}$ in the paramagnetic volume \cite{Dalmas}. Therefore, we have fixed $\beta$ at 1.0 for $\it{T}$ $\leq$ $T_\mathrm{m}$. The PM asymmetry in the sample begins to decrease for $\it{T}$ $\leq$ $T_\mathrm{m}$ due to the growth of the emergent spin clusters described by the $A_\mathrm{E}$ term in Eq. \ref{ZF-HT} (Fig. \ref{fig:3}b $\&$ Fig. \ref{fig:4}b). Because there is no magnetic long-range order at these temperatures \cite{Shen}, our observations unambiguously show the presence of magnetic microphase separation in this material. The paramagnetic fluctuation rate diverges while approaching 6.6 K, below which a coherent oscillation is observed (Fig.~\ref{fig:1}a $\&$ Fig.~\ref{fig:3}c, d), signifying a magnetic phase transition at this temperature. The effective volume of the phase responsible for the coherent (incoherent) muon precession described by the $A_\mathrm{1}$ ($A_\mathrm{2}$) terms in Eq. \ref{ZF-LT} is 69(3)\,$\%$ (see Appendix B). This value is too large to be the $k_\mathrm{2}$ phase, which is a minority in the ground state \cite{Shen}. As a result, we attribute this coherent oscillation to the quasistatic magnetic field generated in the majority $k_\mathrm{1}$ phase, the N\'eel temperature of which ($T_\mathrm{N1}$) is exactly 6.6 K \cite{Shen}. We also note that the small value of $A_\mathrm{1}$ indicates that there is more than one type of muon precession site associated with the $k_\mathrm{1}$ phase (Fig. \ref{fig:3}a). Combining this with the fact that only the muon relaxation rate describing the PM volume diverges at $T_\mathrm{N1}$ (Fig. \ref{fig:3}c $\&$ Fig. \ref{fig:4}c), we conclude that the spins in the PM volume condense into the $k_\mathrm{1}$ state at $T_\mathrm{N1}$ = 6.6 K.

The temperature dependence of the muon relaxation rate describing the emergent spin clusters ($\lambda_{\mathrm{E}}$) is shown in Fig.~\ref{fig:4}c. Unlike the paramagnetic spins, the magnetic fluctuations generated by these short-range correlated clusters diverge at a higher temperature: 7.5 K. Concomitantly, $A_\mathrm{E}$ and $A_\mathrm{NM}$ are suppressed (Fig.~\ref{fig:4}b $\&$ f). These features suggest the existence of a second magnetic phase transition at 7.5 K. The parameters describing these clusters evolve smoothly while cooling below $T_\mathrm{N1}$. This implies a loose coupling between them and the spins in the $k_\mathrm{1}$ domain. $A_\mathrm{NM}$ has a non-vanishing contribution from the sample below $T_\mathrm{m}$, which only drops to the silver background line, $A_\mathrm{NM}$ (Ag) = 0.0420(3), at 5.6 K. This strongly supports the argument that these emergent clusters are intimately coupled to the $k_\mathrm{2}$ phase, which undergoes an incommensurate-commensurate lock-in transition at $T^{\star}$ = 5.6 K \cite{Shen}. Accordingly, the first suppression of $A_\mathrm{NM}$ at 7.5 K, correlated with the divergence of $\lambda_{\mathrm{E}}$, marks the onset of the $k_\mathrm{2}$ phase. Additional support for this comes from the magnetic susceptibility versus temperature measurements, which show a broad peak centered at 7.5 K in our sample (Fig. \ref{fig:S2}a in Appendix C). All these can be explained by the prevailing local FM clusters in the $k_\mathrm{2}$ volume \cite{Kimber,Shen}, the net moment of which only get cancelled out below the N\'eel phase transition at 7.5 K. In other words, we have demonstrated that the N\'eel temperature of the $k_\mathrm{2}$ phase is $T_\mathrm{N2}$ = 7.5 K. 

We now discuss the muon relaxation process described by the $A_{\mathrm{3}}$ term in Eq. \ref{ZF-LT} (Fig. \ref{fig:4}a $\&$ c). First of all, it is not related to the $k_\mathrm{1}$ phase, which has already been captured by the first two terms in Eq. \ref{ZF-LT}. Moreover, its initial asymmetry is too large to account for the domain walls between the two phases. We therefore assign it to the magnetic fluctuations in the $k_\mathrm{2}$ phase. The two magnetic phase transitions in this compound complete around 4.6 K (see Fig. \ref{fig:3}c and Ref.~\onlinecite{Shen}). Upon further cooling, we see that a large portion of the sample volume accommodating the $A_{\mathrm{3}}$ fluctuations becomes NM; this conversion finishes around 2.8 K (Fig. \ref{fig:4}a $\&$ e). The NM effective volume in the sample at 1.8 K, after subtracting the silver holder contribution, is 13(2)\,$\%$. This is close to half of the estimated effective volume of the $k_\mathrm{2}$ phase, which is 30(3)\,$\%$ (Appendix B). Looking at its magnetic structure, half of the spin chains are completely disordered in the $k_\mathrm{2}$ ground state \cite{Shen}. Notably, at the disordered spin site, the four nearest neighbours, which are all magnetically ordered, generate a compensated magnetic field. As a result, we propose that the reentrant NM volume comes from these disordered spins. The residual $A_{\mathrm{3}}$ term below 2.8 K (Fig. \ref{fig:4}a), on the other hand, could come from the weak short-range spin correlations that persist down to at least 1.5 K, as revealed by NPD \cite{Shen}. Accordingly, the non-zero muon decay above 4.6 K is related to the strong spin fluctuations in the transition region or the incommensurate spin arrangement above $T^{\star}$, which can generate a non-compensated magnetic field at the muon stopping sites. These results highlight the metastable nature of the $k_\mathrm{2}$ phase in this compound at intermediate temperatures, which only gets fully stabilized below 2.8 K.

\section{V. Discussion and Summary}

\begin{figure}[b]
	\centering
	\includegraphics[width=0.4\textwidth]{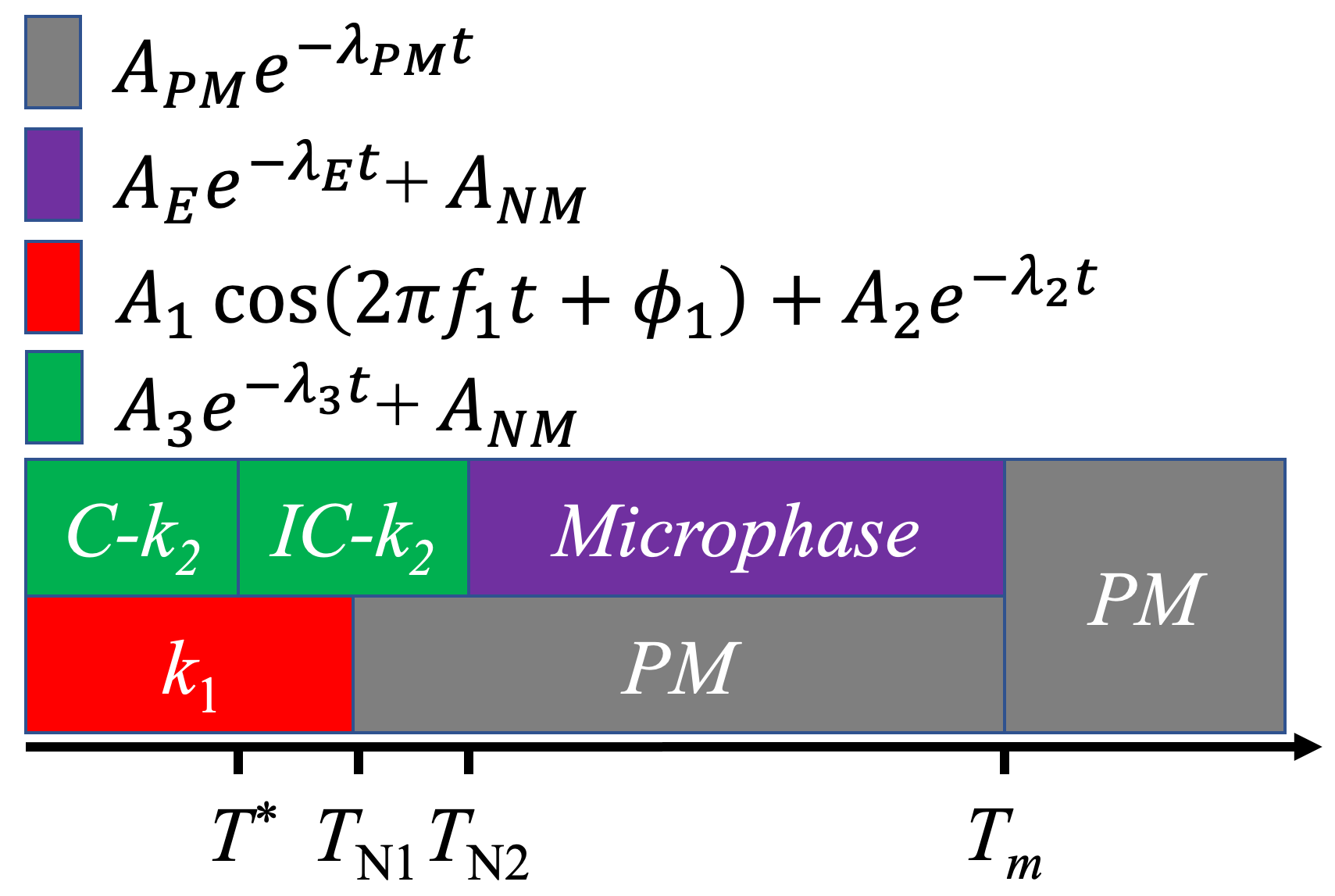}
	\caption{The magnetic phase diagram of $\gamma$CVO as a function of temperature. C, IC, PM correspond to commensurate, incommensurate and paramagnetic, respectively. The term(s) in Eq. \ref{ZF-HT} or \ref{ZF-LT} describing the corresponding phase is (are) also displayed.}
	\label{fig:5}
\end{figure}

With all the results presented above, we discuss the microscopic picture of the magnetic phase separation in $\gamma$CVO. Although no magnetic order is formed above $T_\mathrm{{N2}}$, this system does not fit the Curie-Weiss description between 100 K and $T_\mathrm{m}$ \cite{Kimber}. Because of the positive Weiss temperature in our sample ($\sim$ 14.5 K in our sample, see Appendix C) \cite{Kimber}, which is predominantly contributed by the intrachain coupling, we believe that local spin-spin correlations develop along the chains in this region. From our $\mu$SR data, $\gamma$CVO is magnetically homogeneous above $T_\mathrm{m}$ (Fig. \ref{fig:2}), meaning that the 1D FM clusters should populate the entire sample volume. The magnetic microphase separation sets in when $\gamma$CVO is cooled below $T_\mathrm{m}$.  If we assume that the 1D character persists on cooling, it may be that the effectively isolated spin chains become magnetically inhomogeneous, with no correlations between them, or in the $ac$ plane. Another possibility is that a local 1D-3D dimensional crossover occurs at $T_\mathrm{m}$. In this scenario, the magnetic inhomogeneities described by the $A_\mathrm{{E}}$ term in Eq. \ref{ZF-HT} are exclusively introduced by the energetically relevant interchain coupling(s). Although $\mu$SR is a local probe and therefore cannot directly distinguish these scenarios alone, the higher N\'eel temperature for the spins in these emergent clusters ($T_\mathrm{{N2}}$) seems to favor the second scenario from an energy point of view. This interchain scenario can also explain the puzzle of dominant magnetic diffuse scattering around the $k_\mathrm{2}$ modulated positions below $T_\mathrm{m}$ though it is a minority phase in the ground state \cite{Shen}. While the spins in the emergent clusters ($k_\mathrm{2}$ domain) develop some coherence at $T_\mathrm{N2}$, the true magnetic long-range order, together with its modulation vector, is not stabilized until $T^{\star}$. These features turn out to be strongly correlated with the additional NM muon stopping sites in the sample. As a result, this NM environment must be responsible for the metastable $k_\mathrm{2}$ phase, i.e. its temperature dependent modulation vector and finite spin-spin correlation length \cite{Shen}, between $T_\mathrm{N2}$ and $T^{\star}$. Although the $k_\mathrm{2}$ magnetic phase transition is completed around 4.6 K, strong magnetic fluctuations, described by the $A_\mathrm{{3}}$ term in Eq. \ref{ZF-LT}, can be observed down to 2.8 K. This observation could explain the strong low-energy excitations above 2.0 K observed by INS \cite{Kimber}. As for the $k_\mathrm{1}$ phase, it results from the non-vanishing PM volume in the magnetic microphase separation state. Since no interchain spin correlation exists in these domains, this state is energetically less favored, corresponding to a lower transition temperature $T_\mathrm{N1}$. Below $T_\mathrm{N1}$, the spins in this phase behave like those in a conventional magnet \cite{Dalmas}.

In summary, we have used $\mu$SR, which is an extremely sensitive probe for local magnetic environment \cite{Dalmas}, to follow the development and evolution of the spatially segregate magnetic phases in $\gamma$CVO as a function of temperature. The obtained magnetic phase diagram, along with the $\mu$SR term(s) describing the corresponding phase, is displayed in Fig. \ref{fig:5}. The key finding is the magnetic microphase inhomogeneities that emerge in the PM state at $T_\mathrm{m}$, which we then demonstrated to be a thermodynamic precursor for the ground state phase separation in this material. In the family of quasi 1D magnets, the static and dynamic magnetic properties in the homogeneous state have been intensively studied \cite{Coldea,Bera1,Dupont,Wiersche}. The physics of magnetic phase separation, however, are not well understood in materials of this class and have mostly been focused on materials with a higher effective dimension, e.g. the two-dimensional Kagom\'e or triangular lattice \cite{Lawes,Agrestini,Kamiya} and three-dimensional perovskite lattice \cite{Dagotto}. Regardless of the dimension, one key component shared by all these materials is the strong electron correlation effect, which leads to a situation dominated by frustrated quantum many-body interactions, in which the system's total free energy cannot be minimized by optimizing the interaction energy between every pair of spins or electrons. Our work has unveiled the thermodynamic pathway to the phase separated magnetic ground state in a quasi 1D magnet and therefore provides a concrete foundation for future theoretical and experimental studies.

\section{Acknowledgements}
We thank L. Folkers for kind assistance in the sample preparation stage. We thank Thomas Greber for giving us access to the SQUID. We also gratefully acknowledge the Science and Technology Facilities Council (STFC) for access to muon beamtime at ISIS. The $\mu$SR data collected at ISIS are available at https://doi.org/10.5286/ISIS.E.RB1910224.

\appendix

\section{Appendix A: ROOM-TEMPERATURE X-RAY DIFFRACTION}

\begin{figure}[H]
	\centering
	\includegraphics[width=0.495\textwidth]{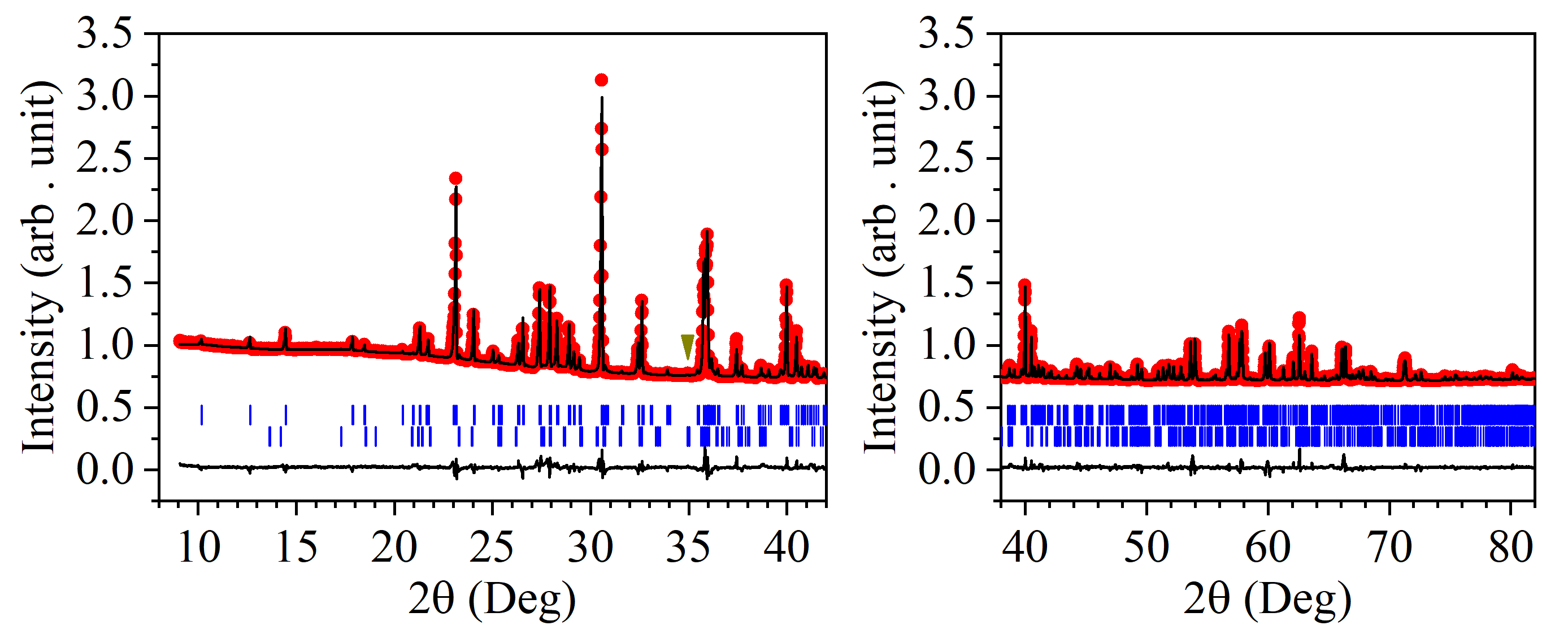}
	\caption{The XRD patterns of $\gamma$CVO at room temperature. The red solid dots are experimental observations. The black lines are the calculated pattern (upper) and the difference between the experimental and calculated data (bottom). The blue vertical bars are the Bragg positions of $\gamma$CVO (upper) and Co$_{2}$V$_{2}$O$_{7}$ (bottom). The dark yellow arrow marks one weak Bragg reflection from the impurity Co$_{2}$V$_{2}$O$_{7}$ phase.}
	\label{fig:S1}
\end{figure}

High-resolution X-ray diffraction (XRD) measurements on our sample were performed at room temperature. The obtained pattern was refined using the Rietveld method in the FullProf package \cite{FullProf}. As shown in Figure \ref{fig:S1}, majority of the Bragg peaks belong to $\gamma$CVO with a volume fraction of 99.55(84) $\%$; the refined lattice parameters and atomic positions are listed in Table \ref{Tab:S1}. These values are in broad agreement with the ones reported in Ref.~\onlinecite{Shen}. In addition to $\gamma$CVO, a very weak impurity phase with a volume fraction of 0.45(18) $\%$, identified as Co$_{2}$V$_{2}$O$_{7}$ \cite{He}, could also be resolved (Fig. \ref{fig:S1}).

\begin{table}[H]
  \centering
  \caption{The refined atomic positions of $\gamma$CVO. The lattice parameters are $\it{a}$ = 7.1799(1) \AA, $\it{b}$ = 8.8992(1) \AA, $\it{c}$ = 4.81306(7) \AA, $\alpha$ = 90.2744(6)$^{\circ}$, $\beta$ = 93.6732(5)$^{\circ}$ and $\gamma$ = 102.1940(6)$^{\circ}$, respectively.}
  \label{Tab:S1}
  \begin{ruledtabular}
  \begin{tabular}{c|ccc}
Atom&$\it{x}$&$\it{y}$&$\it{z}$\\
  \hline
  Co (1)&0&0.5&0\\
  Co (2)&0.0213(5)&0.1715(4)&0.0190(7)\\
  O (1)&0.185(1)&0.503(1)&0.346(2)\\
  O (2)&0.838(1)&0.627(1)&0.143(2)\\
  O (3)&0.177(1)&0.711(1)&0.865(2)\\
  O (4)&0.158(1)&0.031(1)&0.822(2)\\
  O (5)&0.166(1)&0.880(1)&0.335(2)\\
  O (6)&0.791(1)&0.791(1)&0.651(2)\\
  O (7)&0.479(1)&0.942(1)&0.684(2)\\
  O (8)&0.477(1)&0.571(1)&0.700(2)\\
  O (9)&0.524(1)&0.753(1)&0.190(2)\\
  V (1)&0.7178(5)&0.9690(5)&0.4614(8)\\
  V (2)&0.7156(5)&0.6114(5)&0.4551(8)\\
  V (3)&0.5936(6)&0.2592(6)&0.1208(7)\\
  \end{tabular}
  \end{ruledtabular}
\end{table}

\begin{figure*}[t]
	\centering
	\includegraphics[width=0.9\textwidth]{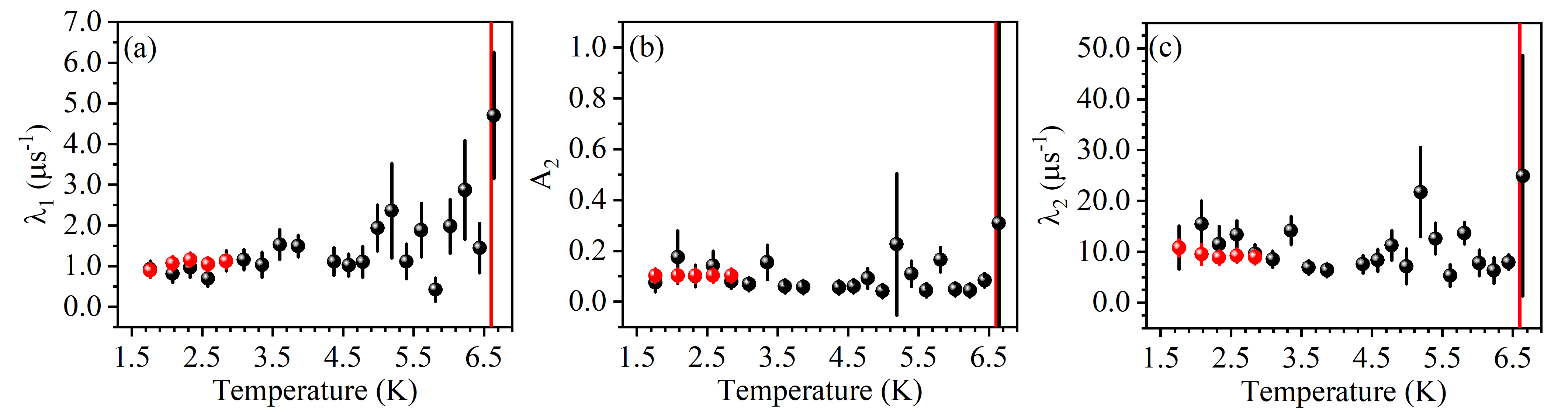}
	\caption{Temperature dependences of (a) $\lambda_\mathrm{1}$, (b) $A_\mathrm{2}$ and (c) $\lambda_\mathrm{2}$ in Eq. 2. The red solid circles are fits based on the additional constraint presented in Section II. The red vertical line marks the position of $T_\mathrm{N1}$ = 6.6 K.}
	\label{fig:S3}
\end{figure*}

\section{Appendix B: Muon spin relaxation and rotation}

The parameters $\lambda_\mathrm{1}$, $A_\mathrm{2}$ and $\lambda_\mathrm{2}$ in Eq. 2 of the main text are plotted in Figure \ref{fig:S3}. The $A_\mathrm{2}$ term captures a fast muon relaxation process; it is only detected in the magnetically ordered region below $T_\mathrm{N1}$ and is typically associated with the incoherent muon precession about the magnetic fields perpendicular to its spin polarization \cite{Monteiro}. Based on the powder average, the effective volume of the muon stopping environment responsible for this fast relaxation is $V_\mathrm{1}$ =  $\dfrac{1.5 \times A_\mathrm{2}}{A_\mathrm{Tot}}$ \cite{Monteiro}; $A_\mathrm{Tot}$ = 0.2224(3) is the total asymmetry of the sample and has been determined from the measurements above $T_\mathrm{m}$ (see the main text). To better determine this value, we performed additional fitting to the $\mu$SR spectra below 2.8 K. Here, we have assumed that $A_\mathrm{1}$ and $A_\mathrm{2}$ in Eq. 2 are temperature independent because the magnetic structure is fully stabilized in this region (see the discussion in the main text). As demonstrated in Fig. \ref{fig:S3}, this approximation significantly reduces the errors of the parameters. $V_\mathrm{1}$ is 69(3)\,$\%$. Based on this value, the magnetic fluctuations described by the $A_\mathrm{2}$ term must come from the $k_\mathrm{1}$ phase because it is the majority \cite{Shen}. Correspondingly, the effective volume of the $k_\mathrm{2}$ phase is 30(3)\,$\%$.

\section{Appendix C: Magnetization versus Temperature}

\begin{figure}
	\centering
	\includegraphics[width=0.495\textwidth]{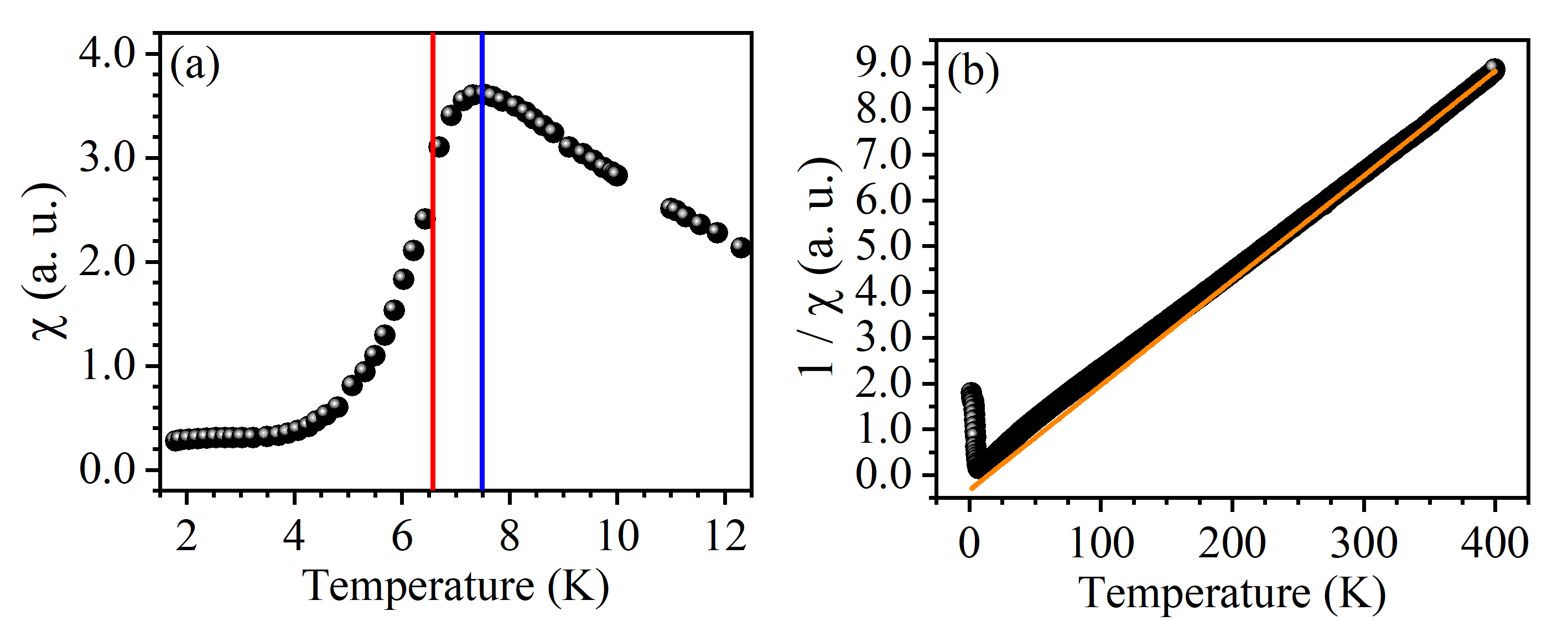}
	\caption{(a) Magnetic susceptibility ($\chi$) as a function of temperature of $\gamma$CVO measured at 0.01 T and below 12.5 K. The red and blue solid lines mark the positions of $T_\mathrm{N1}$ = 6.6 K and $T_\mathrm{N2}$ = 7.5 K, respectively. (b) Inverse $\chi$ versus temperature curve up to 400 K. The orange solid line is the Curie-Weiss fit based on the data points between 350 K and 400 K.}
	\label{fig:S2}
\end{figure}

The magnetic susceptibility ($\chi$) of $\gamma$CVO has been measured as a function of temperature at 0.01 T. As shown in Fig. \ref{fig:S2}a, the broad peak in $\chi$ is centered at 7.5 K; this value matches $T_\mathrm{N2}$ extracted from the $\mu$SR measurements (see the discussion in the main text). We have also plotted out the 1/$\chi$ versus temperature curve in Fig. \ref{fig:S2}b. A Curie-Weiss (CW) fit has been performed on the data points between 350 K and 400 K, which produces a positive Weiss temperature of about 14.5 K. This agrees with the dominant ferromagnetic intrachain spin exchange interactions in $\gamma$CVO \cite{Kimber}. Moreover, a deviation from the CW behaviour is evident above 250 K.

\end{document}